\documentstyle[12pt,epsf]{article}

\font\tenrm=cmr10
\font\tenit=cmti10
\font\elevenbf=cmbx10 scaled\magstep 1
\font\elevenrm=cmr10 scaled\magstep 1
\font\elevenit=cmti10 scaled\magstep 1

\font\ninerm=cmr9

\renewenvironment{thebibliography}[1]
{ \elevenrm
\begin{list}{\arabic{enumi}.}
{\usecounter{enumi} \setlength{\parsep}{0pt}
\setlength{\itemsep}{3pt} \settowidth{\labelwidth}{#1.}
\sloppy
}}{\end{list}}

\begin{document}
\begin{center}
\vglue 0.6cm
{{\elevenbf\vglue 10pt
ELECTROWEAK SYMMETRY BREAKING AT THE\\
\vglue 3pt
SUPERCOLLIDER\\}
\vglue 1.0cm
{\tenrm JONATHAN A. BAGGER \\}
\baselineskip=13pt
{\tenit Department of Physics and Astronomy\\}
\baselineskip=12pt
{\tenit The Johns Hopkins University\\}
\baselineskip=12pt
{\tenit Baltimore, MD\ 21218\\}}
\end{center}

\newcommand{\sm}{SU(3) $\times$ SU(2) $\times$ U(1)}
\newcommand{\Sigmad}{\Sigma^\dagger}
\newcommand{\Tr}{{\rm Tr\,}}
\catcode`\@=11 
\newcommand{\lsim}{\mathrel{\mathpalette\@versim<}}
\newcommand{\gsim}{\mathrel{\mathpalette\@versim>}}
\def\@versim#1#2{\vcenter{\offinterlineskip
\ialign{$\m@th#1\hfil##\hfil$\crcr#2\crcr\sim\crcr } }}
\def\cropen#1{\crcr\noalign{\vskip #1}}
\def\crr{\cropen{1\jot }}
\catcode`\@=12 
\newcommand{\Ref}[1]{(\ref{#1})}
\renewcommand{\L}{{\cal L}}

\vglue 0.5cm
{\elevenbf\noindent 1. Introduction}
\vglue 0.4cm
\baselineskip=14pt
\elevenrm

As we have heard throughout this conference, the standard model
of particle physics is in excellent agreement with experiment.
Precision measurements have confirmed that the strong, weak and
electromagnetic forces are described by a nonabelian gauge theory
based on the group \sm.  This is a remarkable achievement, one for
which particle physicists have every right to feel proud.

To date, however, there has been no experimental evidence in favor
of the Higgs boson $H$.  The Higgs is a central feature of the
standard model because its vacuum expectation value $v$ gives mass
to bosons and fermions alike.  On general grounds, we know that the
Higgs must appear at SSC energies \cite{lqt,cg}.  But for now, we are
frustrated by the fact that all experiments can be described by a
Higgs-free standard model.

The present situation can be summarized by the Lagrangian
\begin{equation}
\L\ =\ \L_{SM}\ +\ \L_{SB}
\end{equation}
where $\L_{SM}$ denotes the gauge part of the standard model and
$\L_{SB}$ the source of electroweak symmetry breaking.  In the
standard model, $\L_{SB}$ contains the Higgs boson $H$, but of
course, it could always contain something else -- new physics
beyond the standard model.  In this talk we will study several
possibilities for $\L_{SB}$ and assess the reach of the
SSC for discovering the physics of electroweak symmetry breaking.

Although present experiments have little to say about $\L_{SB}$,
there are a few basic facts that we will use to guide our discussion:

\begin{itemize}

\item The $W^\pm$ and $Z$ have nonzero masses.  This implies that
$\L_{SB}$ contains at least three would-be Goldstone bosons.  The
Goldstone bosons arise from breaking a group $G$ down to a subgroup
$H$.

\item $M_W = M_Z \cos\theta$.  This tells us that the group $G$
contains SU(2) $\times$ SU(2), and that $H$ contains SU(2).  For the
purposes of this talk, we will call this SU(2)  ``isospin.''  The
would-be Goldstone bosons form a triplet under this isospin symmetry.

\item $\L_{SB}$ is described by a relativistic quantum field
theory.  This provides a framework for discussing alternatives to
the standard model.  In this talk we will assume that $\L_{SB}$ is
described by a strongly-coupled field theory, such as might arise from
a technicolor model.

\end{itemize}

Therefore, in the spirit of technicolor models, we will assume that
the only new particles below 1 TeV are the Goldstone and pseudo-Goldstone
bosons associated with breaking the symmetry group $G$ down to $H$.  The
couplings of these particles  are determined by a gauged chiral Lagrangian,
which provides a model-independent description of their interactions
\cite{ccwz}.  The analysis presented here represents a potential
``worst-case'' scenario for electroweak symmetry breaking at the SSC.

\vglue 0.5cm
{\elevenbf\noindent 2. Gauged Chiral Lagrangian}
\vglue 0.4cm
\elevenrm

We will now construct the gauged chiral Lagrangian that describes the
electroweak symmetry-breaking sector.  The chiral Lagrangian has the great
advantage that it gives a consistent and calculable framework for studying
new physics beyond the standard model.  It correctly incorporates the chiral
symmetries associated with the group $G$ and clarifies the limits of validity
for this approach to symmetry breaking \cite{wwcpt}.

For simplicity, we assume the minimal number of Goldstone particles and take
$G =$ SU(2) $\times$ SU(2) and $H =$ SU(2).  We then introduce the Goldstone
fields $w^a$ and the gauge fields $W^a$ and $B$ through the matrices
\begin{eqnarray}
\Sigma & \equiv & \exp{\bigg( {2 i w^a \tau^a \over v}
\bigg)} \nonumber \\[1mm]
W_\mu & \equiv&  W^a_\mu \tau^a \nonumber \\[1mm]
B_{\mu \nu} & \equiv & (\partial_\mu B_\nu
-\partial_\nu B_\mu ) \tau^3 \nonumber \\[1mm]
W_{\mu \nu} & \equiv & \partial_\mu W_\nu -
\partial_\nu W_\mu + i\,g [W_\mu,W_\nu]\ ,
\end{eqnarray}
where the $\tau^a$ are Pauli matrices, normalized so that $\Tr(\tau^a \tau^b )
= \delta^{ab}/2$.  The derivative
\begin{equation}
D_\mu \Sigma \ =\ \partial_\mu \Sigma\ +\ i\,g W_\mu
\Sigma \ - \ i\,g^{\prime} B_\mu \Sigma \tau^3
\label{cov}
\end{equation}
transforms covariantly under SU(2) $\times$ SU(2),
\begin{equation}
\Sigma\ \rightarrow\  L\, \Sigma\, R^{\dagger}\ ,
\end{equation}
where $L,R \in $ SU(2) and $g, g^{\prime}$ are the coupling constants
of the gauged SU(2)$_L$ and U(1)$_Y$ respectively.

The covariant derivative \Ref{cov} allows us to construct the effective
Lagrangian as a power series in momenta,
\begin{equation}
\L_{SB}\ =\ \L^{(2)}\ +\ \L^{(4)}\ +\ \ldots\ +\ \L^{(res)}\
+\ \ldots\ .
\end{equation}
This is a nonlinear, nonrenormalizable Lagrangian, which must be understood
as a power series in momenta, valid below some scale $\Lambda \lsim 4 \pi v$.
It contains terms $\L^{(n)}$, built out of $n$ powers of the covariant
derivatives $D\Sigma$ and $D\Sigmad$, as well as contributions $\L^{(res)}$
from any explicit TeV-scale resonances.

\begin{figure}[t]
\hspace*{.5in}
\epsfxsize=5.0in\epsffile{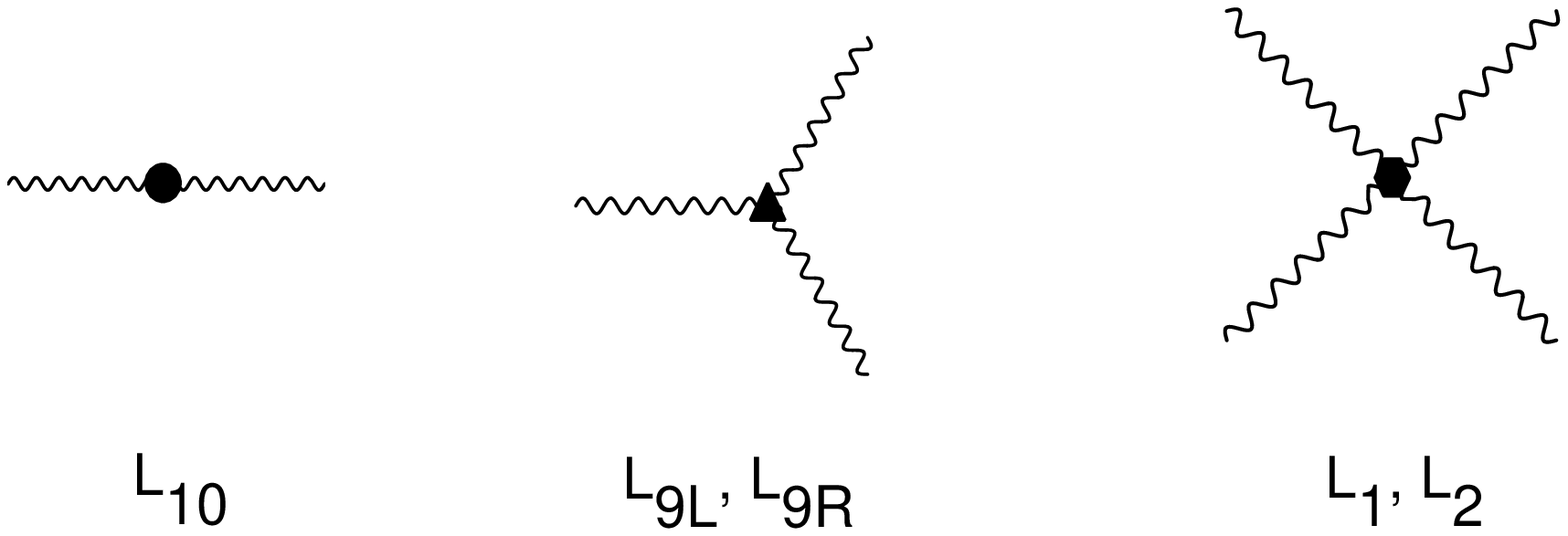}
\begin{quotation}
\noindent
{Fig.~1. The operators in $\L^{(4)}$ give rise to anomalous
gauge-boson\break couplings in the unitary gauge.}
\end{quotation}
\end{figure}

To leading order in the momentum expansion, the Lagrangian $\L^{(2)}$ has no
free parameters:
\begin{equation}
\L^{(2)}\ =\ {v^2 \over 4}\, \Tr
D^{\mu}\Sigma^{\dagger}
D_{\mu}\Sigma\ .
\label{ltwo}
\end{equation}
The couplings of the would-be Goldstone bosons to the SU(2)$_L$ $\times$
U(1)$_Y$
gauge fields are fixed by the covariant derivative.  In unitary gauge, with
$\Sigma = 1$, the Lagrangian \Ref{ltwo} gives rise to mass terms for the
$W^\pm$ and $Z$,
\begin{equation}
\L^{(2)}\ \rightarrow\ {1\over2}\,M^2_Z Z^\mu Z_\mu \ +\ M^2_W W^{+\mu}
W^-_\mu\ ,
\end{equation}
with $M_W = M_Z \cos\theta$.

The next-to-leading order terms in the effective Lagrangian count as four
powers
of momentum \cite{gasser}.  They contain five free
parameters:\footnote{\ninerm\baselineskip=11pt
We have normalized the coefficients so they are ${\cal O}(1)$.}
\begin{eqnarray}
\L^{(4)}& = & L_1 \, \left({v^2 \over \Lambda^2}\right)\, \bigg[
\Tr{D^\mu\Sigmad D_\mu \Sigma} \bigg]^2
\ +\ L_2 \, \left({v^2 \over \Lambda^2}\right)\, \bigg[\Tr{D_\mu\Sigmad D_\nu
\Sigma}\bigg]^2 \nonumber \\[1mm]
& &-\ i g L_{9L} \, \left({v^2 \over \Lambda^2}\right)\, \Tr{W^{\mu \nu} D_\mu
\Sigma D_\nu \Sigmad}
\ -\ i g^{\prime} L_{9R} \, \left({v^2 \over \Lambda^2}\right)\, \Tr{B^{\mu
\nu}
D_\mu \Sigmad D_\nu\Sigma} \nonumber \\[1mm]
& &+\ g g^{\prime} L_{10} \, \left({v^2 \over \Lambda^2}\right)\,\Tr{\Sigma
B^{\mu \nu}
\Sigmad W_{\mu \nu}}\ .
\label{lfour}
\end{eqnarray}
To this order, they are the only terms when $G =$ SU(2)
$\times$ SU(2) and $H$ = SU(2), broken only by the hypercharge
coupling $g'$.

The terms in $\L^{(4)}$ are most easily interpreted in the unitary gauge.
{}From Fig.~1 we see that $L_{10}$ gives a correction to the gauge-boson
two-point function; it is proportional to the Peskin-Takeuchi parameter
$S$ \cite{peskin}.  Similarly, $L_{9L}$ and $L_{9R}$ give corrections
to the gauge-boson three-point functions, while $L_1$ and $L_2$ give
corrections to the four-gauge-boson couplings.

Therefore we can interpret the $L_i$ as representing the anomalous
gauge-boson couplings that come from integrating out new physics at
the scale $\Lambda$.  For example, $L_1$, $L_2$, $L_{9L}$
and $L_{9R}$ can be obtained by integrating out a TeV-scale techni-rho,
while $L_1$ is induced by a heavy Higgs boson $H$ \cite{valencia}.

\begin{figure}[t]
\hspace*{.75in}
\epsfxsize=4.4in\epsffile{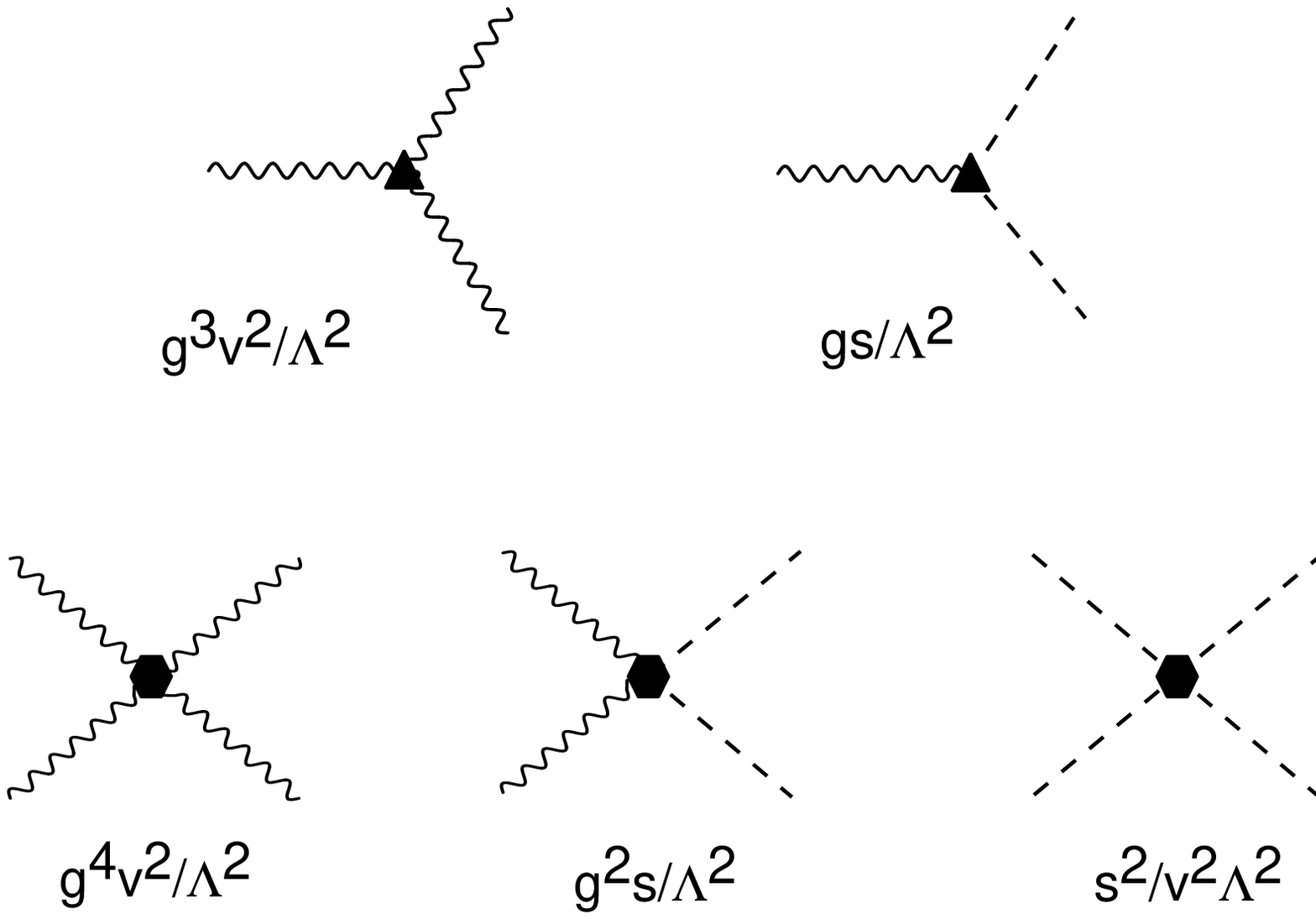}
\begin{quotation}
\noindent
{Fig.~2. The diagrams of enhanced electroweak strength are amplified
by factors of $s/M^2_W$, where $s = E^2$ and $M_W = gv/2$.
The dashed lines represent longitudinally-polarized vector bosons.}
\end{quotation}
\end{figure}

The Lagrangian \Ref{lfour} is very useful in estimating
the size of anomalous gauge-boson couplings.  These couplings are often
presented in the form \cite{peccei}
\begin{eqnarray}
\L_{eff} &= &\sum_{V = \gamma, Z} i g_{WWV} \ \bigg[\,g_V(W^+_{\mu\nu}
W^{-\mu} - W^-_{\mu\nu}W^{+\mu} \,)V^\nu\ +\
i\kappa_V\,W^+_\mu W^-_\nu V^{\mu\nu} \nonumber \\[2mm]
&& \qquad +\
{i\lambda_V\over M^2_W} W^+_{\rho\mu} W^{-\mu}{}_\nu V^{\nu\rho}
\bigg]\ .
\end{eqnarray}
Comparing with equation \Ref{lfour}, in unitary gauge, we find \cite{newapp}
\begin{eqnarray}
g_Z &\equiv& 1\ +\ \Delta g_Z\ \simeq\ 1\ +\ { e^2 L_{9L}\over s^2 c^2}\,
\left({v^2\over\Lambda^2}\right) \ +\
{ 4 e^2 L_{10}\over c^2 (c^2 - s^2)}\,
\left({v^2\over\Lambda^2}\right)
\nonumber \\[1mm]
g_\gamma &=& 1 \nonumber \\[1mm]
\kappa_Z &\equiv& 1\ +\ \Delta\kappa_Z\ \simeq
\ 1\ +\ { e^2(c^2 L_{9L} - s^2 L_{9R})\over s^2 c^2}\,
\left({v^2\over\Lambda^2}\right) \ +\ {8 e^2  L_{10}\over c^2 - s^2}\,
\left({v^2\over\Lambda^2}\right) \nonumber \\[1mm]
\kappa_\gamma &\equiv& 1\ +\ \Delta\kappa_\gamma\ \simeq\ 1\ +\ { e^2(L_{9L} +
L_{9R}
-4 L_{10})\over s^2}\,
\left({v^2\over\Lambda^2}\right) \nonumber \\[1mm]
\lambda_Z &=& 0 \nonumber \\[1mm]
\lambda_\gamma &=& 0 \ ,
\end{eqnarray}
where
$s = \sin\theta$ and $c = \cos\theta$.  Thus, for $\Lambda \sim 1$ TeV, we
expect $\Delta g_Z$ and $\Delta \kappa_V$ to be of order a few percent.  If we
turn the argument around, we see that it is inconsistent to let $\Delta g_Z$
and $\Delta \kappa_V$ be of order one:  if the variations are of order one,
then $v \sim \Lambda$, and the effective Lagrangian has broken down
\cite{burgess,bdv}.

\begin{figure}[t]\hspace*{.4in}
\epsfysize=1.5in\epsffile{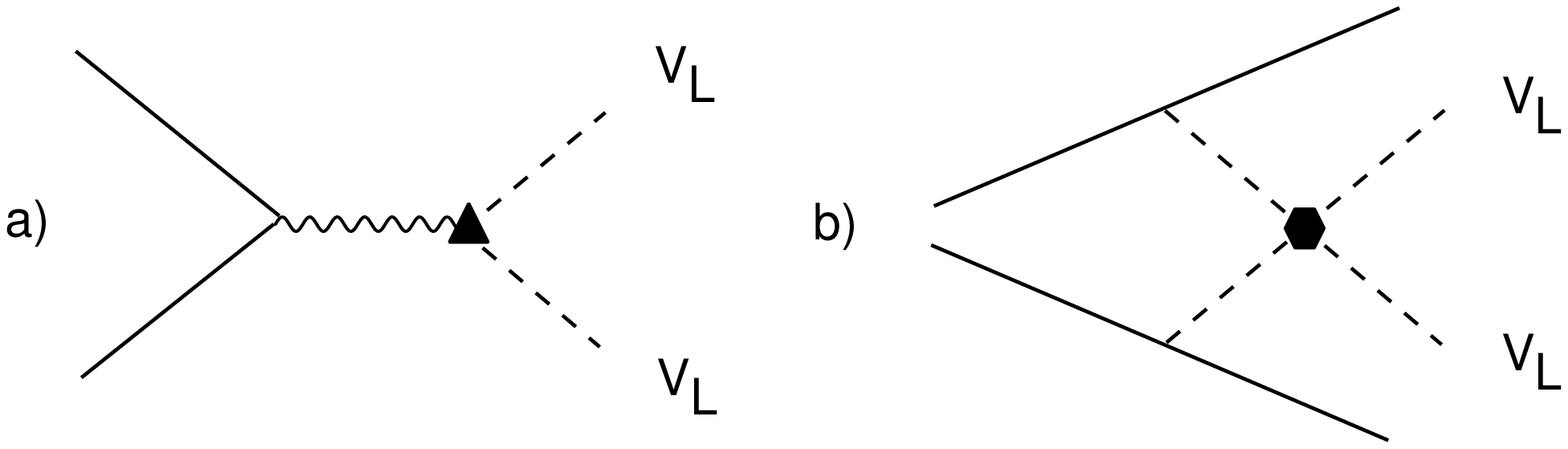}
\begin{quotation}
\noindent
{Fig.~3. Diagrams of enhanced electroweak strength that
contribute to $V_LV_L$ production at hadron supercolliders.}
\end{quotation}
\end{figure}

The effective Lagrangian is also useful in the calculation
of high-energy scattering amplitudes.  According to the electroweak equivalence
theorem \cite{lqt,cg,et},
longitudinally-polarized $W^\pm$'s and $Z$'s can be replaced by
their would-be Goldstone bosons at energies $E \gg M_W$.  Then from the
covariant derivative
\begin{equation}
D_\mu \Sigma\ =\ \partial_\mu \left( {2 i w\over v} \right)\ +\ ig W_\mu\ -\ i
g' B_\mu \tau^3\ \ldots
\end{equation}
we see that scattering amplitudes with transversely-polarized gauge bosons
are suppressed relative to amplitudes with longitudinally-polarized gauge
bosons by factors of $M_W/E \ll 1$.  This implies that the most important
diagrams are those of enhanced electroweak strength, that is, diagrams
with longitudinally-polarized $W^\pm$'s and $Z$'s in loops and external
legs \cite{bdv,eews}.  Other diagrams are less important, as illustrated
in Fig.~2.

Therefore, at hadron supercolliders like the SSC, we expect the best
signals for new physics to be given by the processes
\cite{bdv,falk,others,davis}
\begin{eqnarray}
q \bar q & \rightarrow & W^\pm_L Z_L \nonumber \\
& & W^+_L W^-_L
\end{eqnarray}
and
\begin{eqnarray}
V_L V_L & \rightarrow & W^+_L W^-_L \nonumber \\
& & W^\pm_L Z_L \nonumber \\
& & Z_L Z_L \nonumber \\
& & W^\pm_L W^\pm_L \ ,
\end{eqnarray}
where $V_L = (W^\pm_L,Z_L)$,  as shown in Fig.~3.  The $q \bar q$ initial state
probes $L_{9L}$ and $L_{9R}$ while the $V_L V_L$ initial state is sensitive
to $L_{1}$ and $L_{2}$.
The diagrams of enhanced electroweak strength dominate the search for
electroweak symmetry breaking at high energy supercolliders.\footnote{\ninerm
\baselineskip=11pt Diagrams with photons in the external state are suppressed
by factors of $M_W/E$.  However, they might still be phenomenologically
important because the photon is directly observable, while the useful
modes in $W^\pm$ and $Z$ detection have small branching fractions.}

\vglue 0.5cm
{\elevenbf\noindent 3. Supercollider Signals and Backgrounds}
\vglue 0.4cm
\elevenrm

For the rest of this talk, we will focus on the $V_L V_L$
scattering subprocess shown in Fig.~3b.  This subprocess has the advantage
that it produces final states of all charge combinations.  It is sensitive
to resonant physics of any isospin, as well as to the parameters $L_1$ and
$L_2$.  This subprocess has the additional advantage that it is enhanced by
$(E/M_W)^4$.

The problem with $V_L V_L$ scattering
is that the backgrounds are huge at hadron supercolliders.
For example, there is a large irreducible background from QCD and
standard-model electroweak processes, where the final states are
transversely-polarized $W^\pm$'s and $Z$'s, as shown in Fig.~4.
There is also a large reducible background from top decays.
These backgrounds are much larger than the signal and must be
suppressed by appropriate cuts.

\begin{figure}[t]\hspace*{.25in}
\epsfysize=3.0in\epsffile{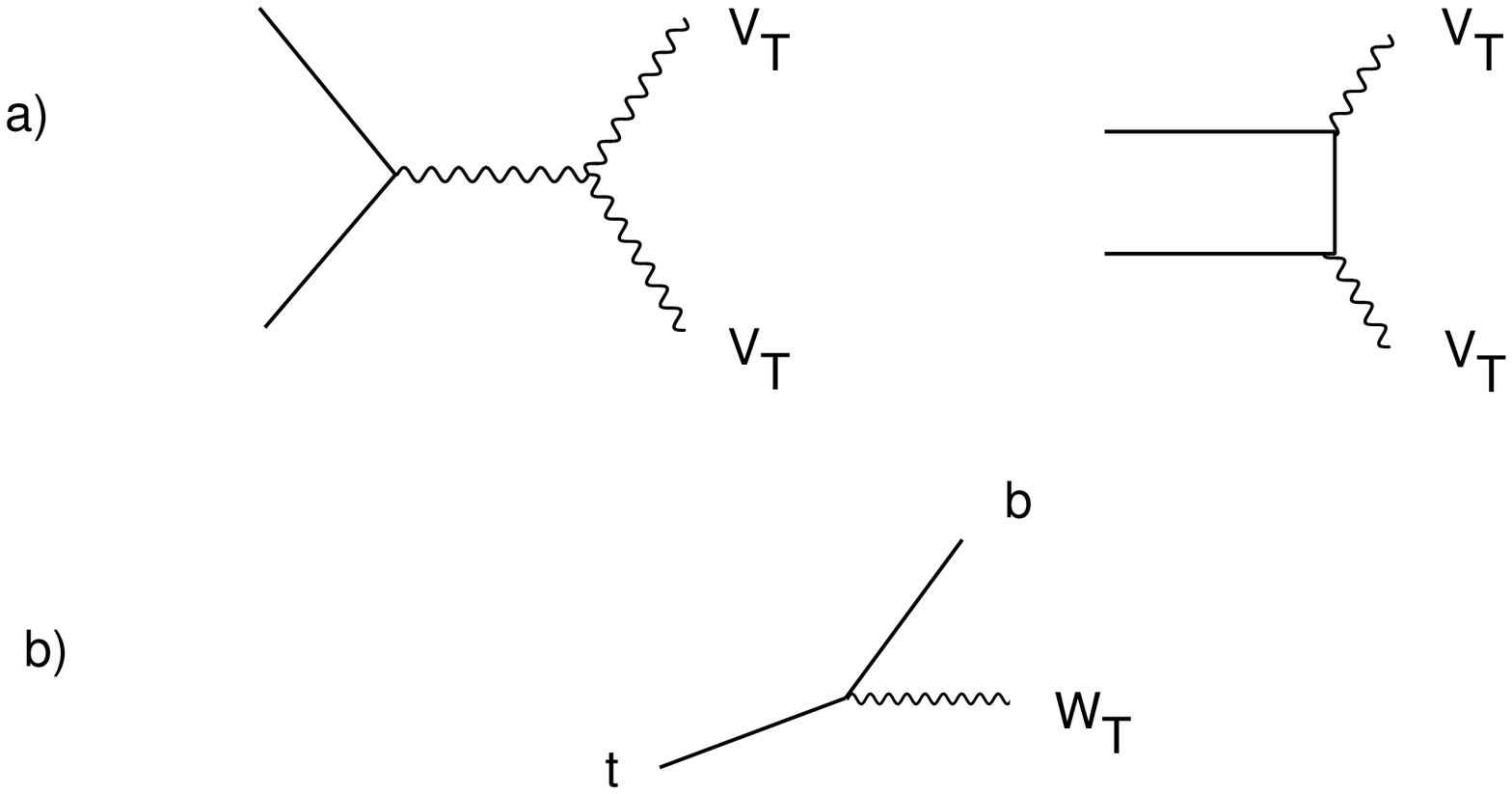}
\begin{quotation}
\noindent
{Fig.~4. The dominant backgrounds are a) irreducible and b)
reducible
production of transversely-polarized $W^\pm$'s and $Z$'s.}
\end{quotation}
\end{figure}

Fortunately, the transversely-polarized background is essentially independent
of $\L_{SB}$.  Therefore it can be computed once and for all using the standard
model with a light Higgs, say $M_H = 100$ GeV.  The result can then be used to
create a set of cuts that suppress the transversely-polarized background
without affecting the longitudinally-polarized signal.

A particularly promising approach was developed in Ref. \cite{davis}.  It has
three parts:

\begin{enumerate}

\item
Focus on the $e$ and $\mu$ final states.  Restricting attention to the
``gold-plated'' final states eliminates many backgrounds that arise when
the $W^\pm$'s and $Z$'s decay to jets.  Imposing
stringent cuts on the leptons, requiring that they be central, energetic,
isolated and back-to-back, suppresses the irreducible transverse background
relative to the longitudinally-polarized signal.

\item
Impose a forward jet tag.  Tagging on
the presence of an energetic, low-$p_T$
jet in the forward calorimeter identifies the jet that radiated the
initial-state $W^\pm$ or $Z$.  This tag reduces the background from $q \bar q$
annihilation in the $W^+W^-$, $W^\pm Z$ and $ZZ$ channels \cite{forward}.

\item
Finally, impose a central jet veto.  This tag suppresses the
reducible
background from top decay.  It is effective in the $W^+W^-$, $W^\pm Z$ and
$W^\pm W^\pm$ channels \cite{central}.

\end{enumerate}

\noindent
The precise set of cuts is listed in Table 1, and is discussed more fully in
K.~Cheung's contribution to these proceedings \cite{cheung}.

\begin{table}[t]
\begin{center}\vspace{-.19in}
\begin{tabular}{ l l | l l}
\multicolumn{4}{l}{\tenrm Table~1. SSC cuts, tags and vetoes, by mode.}\\[2mm]
\hline\hline
$W^+ W^-$ Lepton cuts & Tag and Veto & $Z Z$ Lepton cuts & Tag only\\
\hline
$| y_{\ell} | < 2.0 $  &
$E_{tag} > 1.5\ {\rm TeV}$  &
$| y_{\ell} | < 2.5 $  &
$E_{tag} > 1.0\ {\rm TeV}$   \\
$p_{T,\ell} > 100\ {\rm GeV}$  &
$3.0 < y_{tag} < 5.0$  &
$p_{T,\ell} > 40\ {\rm GeV}$  &
$3.0 < y_{tag} < 5.0$   \\
$\Delta p_{T,\ell\ell} > 450\ {\rm GeV}$  &
$p_{T,tag} > 40\ {\rm GeV}$  &
$p_{T,Z} > {1\over4} \sqrt{M^2_{ZZ} - 4 M^2_Z}$  &
$p_{T,tag} > 40\ {\rm GeV}$   \\
$\cos\phi_{\ell\ell} < -0.8$  &
$p_{T,veto} > 30\ {\rm GeV}$  &
$M_{ZZ} > 500\ {\rm GeV}$  & \\
$M_{\ell\ell} > 250\ {\rm GeV}$  &
$ | y_{veto} | < 3.0$ && \\[2mm]
\hline
$W^+ Z$ Lepton cuts & Tag and Veto &
$W^+ W^+$ Lepton cuts & Veto only \\
\hline
$| y_{\ell} | < 2.5 $  &
$E_{tag} > 2.0\ {\rm TeV}$  &
$| y_{\ell} | < 2.0 $   &
$p_{T,veto} > 60\ {\rm GeV}$  \\
$p_{T,\ell} > 40\ {\rm GeV}$  &
$3.0 < y_{tag} < 5.0$  &
$p_{T,\ell} > 100\ {\rm GeV}$   &
$ | y_{veto} | < 3.0$  \\
$p_{T, miss} >  75\ {\rm GeV}$  &
$p_{T,tag} > 40\ {\rm GeV}$  &
$\Delta p_{T,\ell\ell} > 200\ {\rm GeV}$   & \\
$p_{T,Z} > {1\over4} M_T$ &
$p_{T,veto} > 60\ {\rm GeV}$  &
$\cos\phi_{\ell\ell} < -0.8$  & \\
$M_T > 500\ {\rm GeV}$ &
$ | y_{veto} | < 3.0$  &
$M_{\ell\ell} > 250\ {\rm GeV}$  & \\
\hline\hline
\end{tabular}
\end{center}
\end{table}

To verify that the cuts are effective in isolating the signal from
background, one can compute the signal and background using the standard
model with a 1 TeV Higgs.  Assuming an annual luminosity of 10
fb$^{-1}$ and an energy of $\sqrt{s} = 40$ TeV,  one finds 21 $e,\ \mu$
background events in the $W^+W^-$ channel, 2.5 in the $W^+Z$ channel,
1.0 in the $ZZ$ channel and 3.5 in the $W^+W^+$ channel.  This compares
favorably with the signal in the $W^+W^-$ (36 events) and $ZZ$ (6.8 events)
channels, as expected for an isospin-zero resonance.  (These signals were
computed from the ${\cal O}(\alpha^2)$ matrix elements in the standard
model \cite{davis}.)

\vglue 0.5cm
{\elevenbf\noindent 4. Models for Electroweak Symmetry Breaking}
\vglue 0.4cm
\elevenrm

The above procedure can be used to study electroweak symmetry breaking
for a variety of possible models.  In this section we will discuss
several models that are in accord with all experimental measurements to
date.  Then in the next we will use these models to assess the reach
of the SSC for discovering the origin of electroweak symmetry breaking.

The first major distinction between models is whether or not they are
resonant in the $V_LV_L$ channel.  If a model is resonant, it can be classified
by the spin and isospin of the resonance.  If it is not, it can be described
using the next-to-leading order effective Lagrangian discussed above.  In what
follows, we will restrict our attention to nonresonant models, and to models
with spin-zero, isospin-zero resonances (like the Higgs), and spin-one,
isospin-one resonances (like the techni-rho).

\vglue 0.2cm
{\elevenit \noindent Spin-zero, Isospin-zero Resonances}
\vglue 0.1cm

1) {\elevenit Standard Model.}
The standard model is the prototype of a theory with a spin-zero, isospin-zero
resonance.  The $V_LV_L$ scattering amplitudes are unitarized by exchange of
the Higgs particle $H$.  The Higgs is contained in a complex scalar doublet,
$\Phi\ =\ (v + H) \exp(2i w^a \tau^a/v)$, whose four components
split into a triplet $w^a$ and a singlet $H$ under isospin.

The standard-model Higgs potential is invariant under an SU(2) $\times$ SU(2)
symmetry.  The vacuum expectation value $\langle\Phi\rangle = v$ breaks the
symmetry to the diagonal SU(2).  In the perturbative limit, it also gives
mass to the Higgs.  For the purposes of this talk, we will take $M_H = 1$ TeV.

\vglue 0.1cm
2)  {\elevenit O(2N).}
The second model represents an attempt to describe the
standard-model Higgs in the nonperturbative domain \cite{einhorn}.
In the
perturbatively-coupled standard model, the mass of the Higgs is
proportional to the square root of the scalar self-coupling $\lambda$.
Heavy Higgs particles correspond to large values of $\lambda$.  For $M_H
\gsim$ 1 TeV, naive perturbation theory breaks down, and a
new approach is needed.

One possibility is to exploit the isomorphism between SU(2) $\times$ SU(2)
and O(4).  Using a large-$N$ approximation, one can solve the O(2N) model
for all values of $\lambda$, to leading order in $1/N$.
The resulting scattering amplitudes can be parametrized by the scale $\Lambda$
of the Landau pole.  Large values of $\Lambda$ correspond to small couplings
$\lambda$ and relatively light Higgs particles. In contrast, small values of
$\Lambda$ correspond to large $\lambda$ and describe the nonperturbative
regime.  In this talk we will take $\Lambda = 3$ TeV to represent the
strongly-coupled standard model.

\vglue 0.2cm
{\elevenit \noindent Spin-one, Isospin-one Resonances}
\vglue 0.1cm

1)  {\elevenit Vector.}
This model provides a relatively model-independent description of the
techni-rho resonance that arises in most technicolor theories \cite{bess}.
As above,
one can use nonlinear realizations to construct the effective Lagrangian.
The basic fields are $\xi = \exp(i w^a \tau^a/v)$ and a vector $\rho_\mu$,
which transform as follows under SU(2) $\times$ SU(2),
\begin{eqnarray}
\xi & \rightarrow & L\,\xi\,U^\dagger\ =\ U\,\xi\,R^\dagger\ ,
\nonumber \\
\rho_\mu & \rightarrow & U\rho_\mu\,U^\dagger + i g^{\prime
\prime-1}\, U \partial_\mu U^\dagger\ ,
\end{eqnarray}
where $U(L,R,\xi) \in$ SU(2).

For the processes of interest, the effective Lagrangian depends on just
two couplings, the mass and the width of the resonance.  In what follows
we will choose $M_\rho = 2.0$ TeV, $\Gamma_\rho = 700$ GeV and $M_\rho =
2.5$ TeV, $\Gamma_\rho = 1300$ GeV.  These values preserve unitarity up
to 3 TeV.

\begin{table}[t]
\begin{center}\vspace{-.19in}
\begin{tabular}{ l | c  c  }
\multicolumn{3}{l}{\tenrm Table~2. Efficiencies for tagging and vetoing at the
SSC.}\\[2mm]
\hline\hline
$W^+ W^-$ &  Veto only & Veto plus Tag \\
& 57\% & 38\% \\
\hline
$W^+ Z$   & Veto only & Veto plus Tag \\
& 75\%  & 40\% \\
\hline
$Z Z $    & Tag only & Veto plus Tag \\
& 59\% & $-$ \\
\hline
$W^+ W^+$ &Veto only & Veto plus Tag \\
& 69\% & $-$ \\
\hline\hline
\end{tabular}
\end{center}
\end{table}

\vglue 0.2cm
\noindent
{\elevenit \noindent Nonresonant models}
\vglue 0.1cm

The final models we consider are nonresonant at SSC energies.  In this case
new physics contributes to the effective Lagrangian through the coefficients
$L_i$.  As discussed above, only $L_1$ and $L_2$ contribute to $V_L V_L$
scattering to order $p^4$ in the energy expansion.

The difficulty with this approach is that the scattering
amplitudes violate unitarity between 1 and 2 TeV.  This is an indication that
new physics is near, but there is no guarantee that new resonances lie
within the reach of the SSC.
We choose to treat the uncertainties of unitarization in two ways:

\begin{table}[t]
\begin{center}\vspace{-.19in}
\begin{tabular}{ l | c  c  c  c  c  c  c }
\multicolumn{8}{l}{\tenrm Table~3. Event rates per SSC-year, assuming $m_t
= 140$ GeV, $\sqrt{s} = 40$ TeV,}\\
\multicolumn{8}{l}{\tenrm \hspace{38pt}  and an annual luminosity of
$10$ fb$^{-1}$.}\\[2mm]
\hline\hline
$W^+ W^-$ & Bkgd & SM 1.0 &  O(2N) &  Vec 2.0 & Vec 2.5 & LET CG & Delay K
\\
\hline
$M_{\ell\ell} > 0.25$ & 21  & 48   & 24 &    15  & 12  &
16   & 11  \\
$M_{\ell\ell} > 0.5$  & 17   & 46   & 23 &  15 & 12 &
15 & 11  \\
$M_{\ell\ell} > 1.0$  & 3.6   & 3.8 & 2.7 &  6.5 & 4.9 &
5.3 & 4.6
\\[2mm]
\hline
$W^+ Z$ & Bkgd & SM 1.0 &  O(2N) &  Vec 2.0 & Vec 2.5 & LET CG & Delay K  \\
\hline
$M_T > 0.5$ & 2.5 & 1.3 &  1.5 &  9.5  &  6.2
& 5.8 & 6.0  \\
$M_T > 1.0$ & 0.8 & 0.5 &  0.7 &  7.9  &  4.7
& 4.1 & 4.6  \\
$M_T > 1.5$ & 0.3 & 0.2 &  0.3 &  5.5  &  3.2
& 2.6 & 3.2
\\[2mm]
\hline
$Z Z$ & Bkgd & SM 1.0 &  O(2N) &  Vec 2.0 & Vec 2.5 & LET CG & Delay K  \\
\hline
$M_{ZZ} > 0.5$ & 1.0 & 11  &  5.2 &  1.1 & 1.5 &
2.6 &  1.6
\\
$M_{ZZ} > 1.0$ & 0.3 & 4.1 &  2.0 &  0.4 & 0.7 &
1.6 &  0.8
\\
$M_{ZZ} > 1.5$ & 0.1 & 0.5 &  0.5 &  0.1 & 0.3 &
0.9 &  0.4
\\[2mm]
\hline
$W^+ W^+$ & Bkgd & SM 1.0 &  O(2N) &  Vec 2.0 & Vec 2.5 & LET CG &  Delay K
\\
\hline
$M_{\ell\ell} > 0.25$ & 3.5 & 6.4 &  7.1 &  7.8 & 11 &
25 & 15 \\
$M_{\ell\ell} > 0.5$ & 1.5 & 3.2 &  3.9 &  3.8 & 6.3 &
19 & 11  \\
$M_{\ell\ell} > 1.0$ & 0.2 & 0.7 &  0.9 &  0.5 & 1.2 &
7.6 & 5.2  \\
\hline\hline
\end{tabular}
\end{center}
\end{table}

\vglue 0.1cm
1)  {\it LET CG.}
We take $L_1 = L_2 = 0$, and cut off the partial wave amplitudes when they
saturate the unitarity bound.  This is the original model considered by
Chanowitz and Gaillard \cite{cg,berger}.
\vglue 0.1cm

2)  {\it Delay K.}
We take $L_1 = -0.26$ and $L_2 = 0.23$, a choice that preserves unitarity up to
2 TeV.  Beyond that scale, we unitarize the scattering amplitudes with a
K-matrix \cite{delay}.
\vglue 0.1cm
\noindent
These two models describe possible nonresonant new physics at the SSC.

\begin{figure}[t]
\epsfxsize=6.0in\epsffile{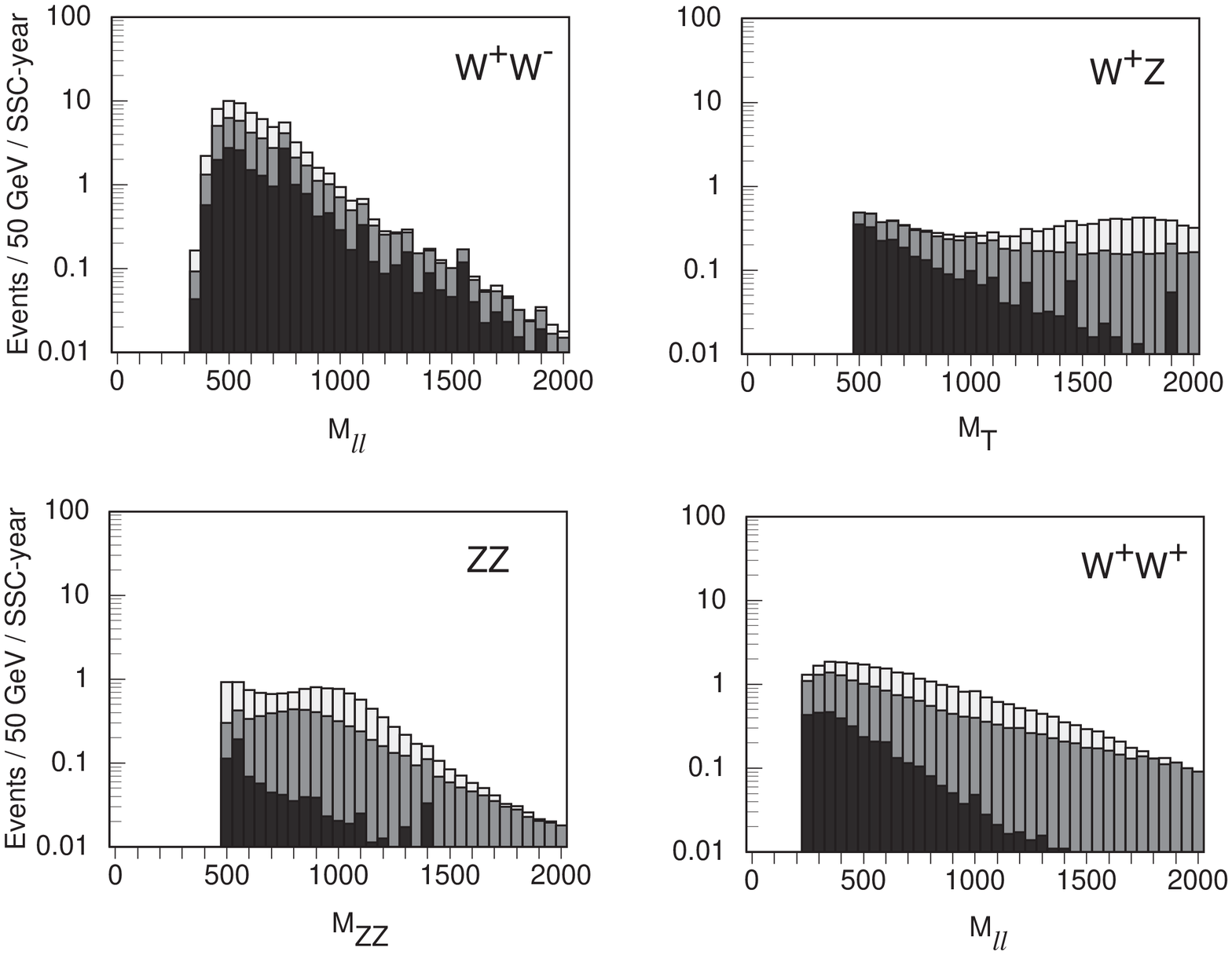}
\begin{quotation}
\noindent
{Fig.~5. Invariant mass distributions for the
$e, \mu$ final states that arise from the processes $pp \rightarrow
W^+W^-X$, $pp \rightarrow W^+Z X$, $pp \rightarrow ZZX$ and
$pp \rightarrow W^+W^+X$, for $\sqrt s = 40$ TeV
and $\int \L dt =10$ fb$^{-1}$.
The darkest histogram always contains the background,
that is, the standard model with a 100 GeV Higgs.  In the $W^+W^-$
and $ZZ$ channels, the medium histogram includes the signal
plus background for the O(2N) model, while the light histogram
contains the signal plus background for the standard model with
a 1 TeV Higgs.  In the $W^+Z$ channel the
light histogram contains the contribution from a vector
isovector resonance with $M_\rho = 2$ TeV and $\Gamma_\rho =
700$ GeV, while
the medium histogram gives the result for $M_\rho = 2.5$ TeV
and $\Gamma_\rho = 1300$ GeV.  The $W^+W^+$ channel includes the
signal plus background from two nonresonant models.  The light
histogram contains the LET CG; the medium, Delay K.}
\end{quotation}
\end{figure}

\vglue 0.5cm
{\elevenbf\noindent 5. Results and Discussion}
\vglue 0.4cm
\elevenrm

\begin{table}[t]
\begin{center}\vspace{-.19in}
\begin{tabular}{ c | c  c  c  c  c  c }
\multicolumn{7}{l}{\tenrm Table~4. Number of SSC-years (if $<10$) required for
a 95\% confidence level signal}\\[2mm]
\hline\hline
Channel $\backslash$ Model & SM 1.0 &  O(2N) &  Vec 2.0 & Vec 2.5 & LET CG &
Delay K\\
\hline
$W^+W^-$ & 0.25  & 0.75  & 1.2   & 1.8   & 1.2  & 2.0  \\
$W^+Z$   & $-$ & $-$ & 0.75  & 1.5   & 1.8  & 1.5  \\
$ZZ$     &1.2    & 3.0   & $-$ & $-$ & 4.0  & $-$   \\
$W^+W^+$ &3.2    & 2.2   & 2.5   & 1.2   & 0.25 & 0.50 \\
\hline\hline
\end{tabular}
\end{center}
\end{table}

For each of the above models, one can compute the signal and background at the
SSC.  As discussed previously, the background is the same for
all models, and can be computed using the full standard model with a 100 GeV
Higgs.  The signal, however, must be computed independently for each model.
The calculation can be simplified by using the effective $W$ approximation
\cite{ewa} and the electroweak equivalence theorem \cite{lqt,cg,et}
to identify the terms of enhanced
electroweak strength \cite{bdv,eews}.  One can then apply efficiencies for
jet tagging and vetoing, derived from the exact standard model
calculation with
a 1 TeV Higgs.  (The efficiencies are collected in Table 2.)  This procedure
has been shown to work for the standard model with a heavy Higgs, and since
the efficiencies depend on the kinematics of the initial state, and not on the
details of the  hard scattering, they should hold equally well for
all the models considered here \cite{davis}.

The final results, including efficiencies and branching fractions into the
$e,\mu$ final states, are listed in Table 3.  The rates are given in
events per SSC-year, assuming $\int \L dt = 10$ fb$^{-1}$.
They are presented as a function of an invariant mass cut on
the final-state observables:  $M_{\ell\ell}$, the dilepton invariant mass in
the $W^+ W^-$ and $W^+ W^+$ channels; $M_{ZZ}$, the $ZZ$ invariant mass in
the $ZZ$ channel; and $M_T$, the cluster transverse mass in the $W^+ Z$
channel.
(Similar results hold for the LHC provided that overlapping events can
be separated and the forward jet tag imposed.)

{}From Table 3 we see that the total event rates are rather low.  As expected,
they are largest in the resonant channels.  The events are clean, but the low
rates make it difficult to hunt for bumps.  This is
emphasized in Fig.~5, where invariant mass distributions are presented
for several of the models \cite{davis}.

To clarify these results, let us define a signal to be observable if the
maximum background at 95\% C.L. is less than the minimum signal plus background
at the same confidence level.  Then one can tabulate the number of years
required to see the signal in each channel.  This is done in Table 4, where
the middle invariant mass cut is used for each mode.  Table 4 shows that each
model is observable in some mode after a few years of SSC operation
\cite{davis}.

The results in Tables 3 and 4 indicate that all channels must be measured if
one is to be sure to see electroweak symmetry breaking at the
SSC. For example, isospin-zero resonances give the best signal in the $W^+W^-$
and $ZZ$ channels, while isospin-one resonances dominate the $W^+ Z$ channel.
The nonresonant models tend to show up in the $W^+W^+$ final state, so there
is a complementarity between the different channels \cite{chandallas}.

Clearly, because of the low rates, one cannot cut corners.  Accurate background
studies are essential if one is to separate signal from background.  One
must try to measure all decay modes of the $W^\pm$ and $Z$, including $Z
\rightarrow\nu\bar\nu$ and $W^\pm,Z \rightarrow jets$, and one must work to
optimize the cuts that are applied to each final state.  Finally, one must
leave
open the possibility of a luminosity upgrade for the SSC.  Nevertheless,
the results presented here indicate that, with a mature long-term
program, SSC experiments should indeed discover the origin of mass.

\vglue 0.5cm
{\elevenbf\noindent 6. Acknowledgments}
\vglue 0.4cm
\elevenrm

I would like to thank V.~Barger, K.~Cheung, S.~Dawson, J.~Gunion, T.~Han,
G.~Ladinsky, R.~Rosenfeld, G.~Valencia and C.-P.~Yuan for pleasant
collaborations on the effective Lagrangian approach to electroweak
symmetry breaking.

\vglue 0.5cm
{\elevenbf\noindent 7. References \hfil}
\vglue 0.4cm

\end{document}